\documentclass[reprint, double column, superscriptaddress,prl, showkeys]{revtex4-1}

\usepackage{float}
\usepackage{graphicx,epsfig}
\usepackage{amssymb}
\usepackage{amsmath}
\usepackage{bm}
\usepackage{braket}

\usepackage{makecell}
\usepackage{textcomp}
\usepackage{color}
\usepackage{pgffor}
\usepackage{pdfpages}

\usepackage{soul}

\usepackage[bookmarksnumbered,bookmarksopen]{hyperref}
\hypersetup{
   colorlinks=true,       % false: boxed links; true: colored links
}

\usepackage{BOONDOX-cal}

\makeatletter

\AtBeginDocument{\let\LS@rot\@undefined}
\makeatother
\newif\ifarXiv
\arXivtrue

\usepackage{soul}

\begin{document}
\setcounter{page}{1}

\title[]{Developing fractional quantum Hall states at even-denominator fillings 1/6 and 1/8}
\author{Chengyu \surname{Wang}}
\author{P. T. \surname{Madathil}}
\author{S. K. \surname{Singh}}
\author{A. \surname{Gupta}}
\author{Y. J. \surname{Chung}}
\author{L. N. \surname{Pfeiffer}}
\author{K. W. \surname{Baldwin}}
\author{M. \surname{Shayegan}}
\affiliation{Department of Electrical and Computer Engineering, Princeton University, Princeton, New Jersey 08544, USA}
\date{\today}

\begin{abstract}
In the extreme quantum limit, when the Landau level filling factor $\nu<1$, the dominant electron-electron interaction in low-disorder two-dimensional electron systems leads to exotic many-body phases. The ground states at even-denominator $\nu=$ 1/2 and 1/4 are typically Fermi seas of composite fermions carrying two and four flux quanta, surrounded by the Jain fractional quantum Hall states (FQHSs) at odd-denominator fillings $\nu=p/(2p\pm1)$ and $\nu=p/(4p\pm1)$, where $p$ is an integer. For $\nu<1/5$, an insulating behavior, which is generally believed to signal the formation of a pinned Wigner crystal, is seen. Our experiments on ultrahigh-quality, dilute, GaAs two-dimensional electron systems reveal developing FQHSs at $\nu=p/(6p\pm1)$ and $\nu=p/(8p\pm1)$, manifested by magnetoresistance minima superimposed on the insulating background. In stark contrast to $\nu=1/2$ and 1/4, however, we observe a pronounced, sharp minimum in magnetoresistance at $\nu=1/6$ and a somewhat weaker minimum at $\nu=1/8$, suggesting developing FQHSs, likely stabilized by the pairing of composite fermions that carry six and eight flux quanta. Our results signal the unexpected entry, in ultrahigh-quality samples, of FQHSs at \textit{even-denominator} fillings 1/6 and 1/8, which are likely to harbor non-Abelian anyon excitations.
\end{abstract}

\maketitle

When a two-dimensional electron system (2DES) at low temperatures is subjected to a large, perpendicular magnetic field ($B$), the electrons’ kinetic energy is quenched as they occupy quantized, dispersionless Landau levels (LLs). The dominant electron-electron Coulomb energy leads to a variety of exotic, strongly-correlated, many-body ground states, depending on the LL filling factor $\nu=nh/eB$, where $n$ is the 2DES density. When $\nu$ is a rational fraction, fractional quantum Hall states (FQHSs), incompressible liquid states that host quasiparticles with fractional charge and anyonic statistics, manifest as the ground states \cite{Tsui.PRL.1982, Jain.Book.2007, Halperin.Book.2020, Nakamura.Natphys.2020, Bartolomei.science.2020}. Of particular interest are FQHSs observed at \textit{even-denominator} fillings of excited-state ($N$=1) LLs, e.g., at $\nu=$5/2 \cite{Willett.PRL.1987}, because they are likely to harbor non-Abelian anyon excitations \cite{Banerjee.Nature.2018, Willett.PRX.2023}, which can be useful for topological quantum computation \cite{Nayak.RMP.2008}.
However, the vast majority of FQHSs are observed in the extreme quantum limit ($\nu<$ 1) at \textit{odd-denominator} fillings, and are successfully explained by Laughlin's wave function \cite{Laughlin.PRL.1983} and Jain's composite fermion (CF) theory \cite{Jain.PRL.1989, Jain.Book.2007}.

Another longstanding, fundamental, and important question relates to the fate of a clean 2DES at extremely small $\nu$. It is generally believed that, for sufficiently small $\nu$, electrons should form an ordered array, known as the Wigner crystal (WC) \cite{Wigner.PR.1934, Shayegan.NatRevPhys.2022}. Theorists predict a termination of the FQHSs and transition into a quantum WC at a critical $\nu$ ranging from 1/6 to 1/11 \cite{Laughlin.PRL.1983, Lam.PRB.1984, Esfarjani.PRB.1990, Yang.PRB.2001, Zuo.PRB.2020}. On the experimental front, strong evidence for pinned WC states was reported at $\nu\lesssim1/5$ in GaAs 2DESs \cite{Andrei.PRL.1988, Jiang.PRL.1990, Goldman.PRL.1990, Chen.NatPhys.2006, Deng.PRL.2016}. The WC exhibits an insulating behavior because of the pinning by the ubiquitous disorder in a realistic (nonideal) 2DES. In the highest quality samples, signatures of FQHSs were also reported in the very small filling regime, e.g. at $\nu=$1/7, in the form of resistance minima superimposed on the strongly insulating background \cite{Goldman.PRL.1988, Pan.PRL.2002, Chung.PRL.2022}. These observations highlight the very close competition between the WC and FQHS phases.

In this work, we examine the regime of $\nu\ll$ 1  in ultrahigh-quality GaAs 2DESs. We observe an unexpected emergence of new correlated states deep in the WC regime, namely \textit{even-denominator} FQHSs at $\nu=$1/6 and 1/8. These states are likely non-Abelian FQHSs stabilized by the pairing of six-flux and eight-flux CFs ($^6$CFs and $^8$CFs). The presence of these large-flux CFs is also evinced by the observation of numerous odd-denominator FQHSs on the flanks of $\nu=$1/6 and 1/8, following the Jain sequence of $^6$CFs and $^8$CFs.

\begin{figure*}[t]
  \begin{center}
    \psfig{file=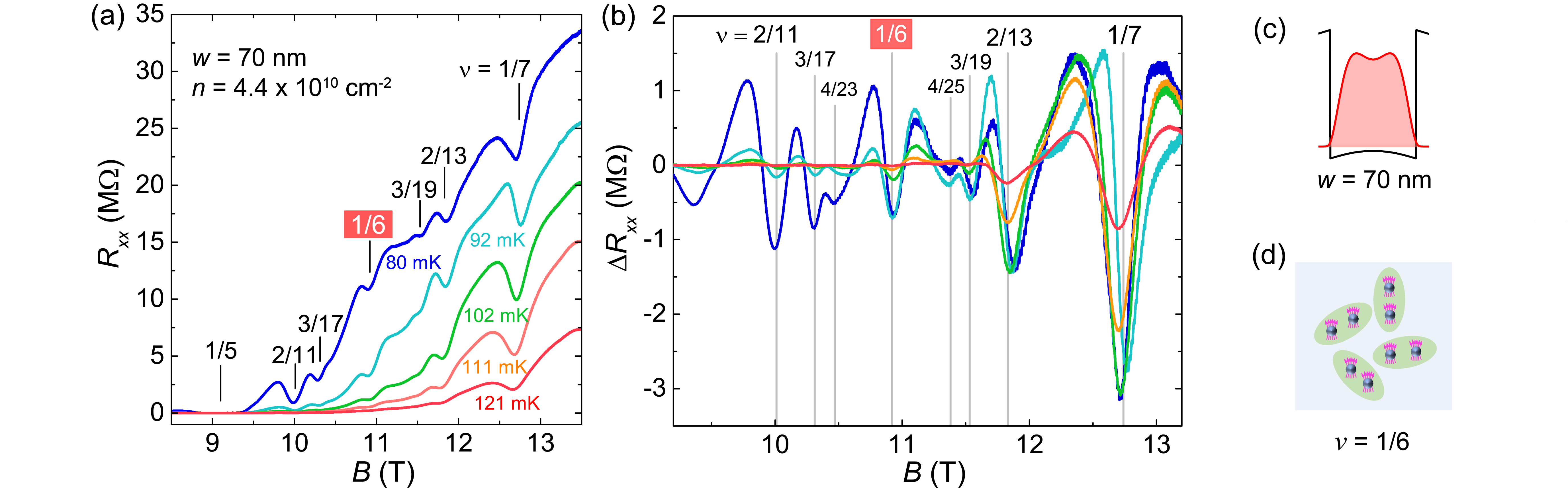, width=1\textwidth}
  \end{center}
  \caption{\label{main}
    (a) Longitudinal resistance ($R_{xx}$) vs perpendicular magnetic field ($B$) traces for our ultrahigh-mobility 2DES in the extremely small filling regime (1/5 $>\nu>$ 1/7), measured at different temperatures \cite{Footnote.lowcurrent}. The 2DES is confined to a 70-nm-wide QW, and has a density of 4.4$\times$10$^{10}$ cm$^{-2}$ and a record mobility of 22$\times$10$^{6}$ cm$^2$/Vs at this density. The magnetic field positions of several LL fillings are marked. Our data exhibit numerous local minima in $R_{xx}$ at odd-denominator fillings 1/5, 2/11, 3/17, 1/7, 2/13, and 3/19. These fillings correspond to the Jain-sequence states of six-flux CFs ($^6$CF). Remarkably, we also observe a local minimum in $R_{xx}$ at the \textit{even-denominator} filling $\nu=$1/6, suggesting a developing FQHS. (b) $\Delta R_{xx}$ vs $B$ traces for the same set of data, where $\Delta R_{xx}$ is the resistance after subtracting the increasing, smooth background \cite{SM}.  (c) Self-consistent charge distribution (red) and potential (black) for the 2DES. (d) A possible origin of the 1/6 FQHS: Each electron captures six flux quanta to turn into a $^6$CF. Then $^6$CFs undergo a pairing instability and condense into a FQHS. 
    }
  \label{fig:main}
\end{figure*}

We studied high-quality 2DESs confined to GaAs quantum wells (QWs) grown on GaAs (001) substrates by molecular beam epitaxy. They were grown following the optimization of the growth chamber vacuum integrity and the purity of the source materials \cite{Chung.Natmater.2021}. We used 4$\times$4 mm$^2$ van der Pauw geometry samples with alloyed In:Sn contacts at the four corners and side midpoints. The samples were cooled in a dilution refrigerator. We measured the longitudinal resistances ($R_{xx}$) using the conventional lock-in amplifier technique.

As highlighted in Fig. \ref{fig:main}(a), on the flanks of $\nu=$1/6, we observe a sequence of minima at $\nu=$1/5, 2/11, 3/17, and 1/7, 2/13, 3/19, superimposed on an extremely large and insulating $R_{xx}$ \cite{Footnote.lowcurrent}. These are the Jain-sequence FQHSs of $^6$CFs [$\nu=p/(6p\pm1)$], emanating from $\nu=$1/6, analogous to the standard Jain-sequence FQHSs of $^2$CFs and $^4$CFs observed on the flanks of CF Fermi seas at $\nu=$1/2 and 1/4; see Fig. S4 in Supplemental Material (SM) \cite{SM} for data near $\nu=$1/2 and 1/4. Our observation is consistent with recent calculations which indicate that, in the low-disorder limit, Jain-sequence FQHSs of $^6$CFs should prevail at $\nu=$1/7 and 2/13 \cite{Zuo.PRB.2020}. The appearances of the very-high-order FQHSs near $\nu=$1/2 and 1/4, and the rarely observed FQHSs near $\nu=$1/6 \cite{Goldman.PRL.1988, Pan.PRL.2002, Chung.PRL.2022}, collectively demonstrate the exceptionally high quality of our 2DES, specially at such a low density ($n=$4.4, in units of 10$^{10}$ cm$^{-2}$, which we use throughout the paper).

Our main finding is the pronounced, sharp minimum in $R_{xx}$ at the \textit{even-denominator} filling $\nu=$1/6. As we illustrate below, the characteristics of this minimum are very similar to those of the nearby, emerging, odd-denominator FQHSs. Our data signal a developing even-denominator FQHS at $\nu=$1/6, likely stabilized by the pairing of $^6$CFs. 

Figure \ref{fig:main}(a) also shows high-field $R_{xx}$ vs $B$ traces at different temperatures. As $T$ increases from 80 to 121 mK, $R_{xx}$ at $\nu<$1/5 decreases by more than an order of magnitude. Meanwhile, $R_{xx}$ minima at $\nu=$1/6 and $\nu=p/(6p\pm1)$ gradually weaken and eventually turn into inflection points \cite{SM, Footnote.activation}. To highlight FQHS features, in Fig. 1(b), we present $\Delta R_{xx}$ vs $B$ traces, with $\Delta R_{xx}$ representing resistance after subtracting the smooth background; see Section I of SM for details \cite{SM}. We observe sharp $\Delta R_{xx}$ minima at $\nu=p/(6p\pm1)$ for $p=$1, 2, 3, 4, and at $\nu=$1/6. The $\nu=p/(6p\pm1)$ minima are weaker for larger $p$ and weaken with increasing $T$, consistent with standard Jain-sequence FQHSs. The $\nu=$1/6 minimum is sharp and exhibits similar temperature dependence to those at Jain-sequence fillings, signaling a developing FQHS at $\nu=$1/6.

We note that with decreasing temperature, instead of approaching zero, $R_{xx}$ at $\nu=$1/6 and $p/(6p\pm1)$ increases. This is because an insulating behavior, which is a manifestation of a pinned WC \cite{Andrei.PRL.1988, Jiang.PRL.1990, Goldman.PRL.1990, Chen.NatPhys.2006, Deng.PRL.2016}, is dominant in the whole range of $\nu<$1/5. Our observation signals a close competition between the FQHSs and WC states. More specifically, the energies of WC and FQHSs are so close that FQHSs only win in a very narrow range of $\nu$ \cite{Zuo.PRB.2020}. Therefore, in a realistic 2DES, a small local variation of filling factor caused by a minuscule density inhomogeneity or disorder can lead to the formation of WC domains and prevent the percolation of the FQH liquid \cite{Footnote.Hall}. Our data are reminiscent of what was historically observed at $\nu=$1/5 in GaAs 2DESs \cite{Mendez.PRB.1983, Willett.PRB.1988, Jiang.PRL.1990, Goldman.PRL.1990, Goldman.PRL.1993, Sajoto.PRL.1993}. Initially, in modest-quality samples, only an $R_{xx}$ minimum that rose with decreasing temperature was seen because of the significant amount of disorder \cite{Mendez.PRB.1983, Willett.PRB.1988}. With improved sample quality, percolation of the FQH liquid was eventually achieved, exhibiting a vanishing $R_{xx}$ accompanied by a quantized Hall plateau, firmly establishing that the ground state at $\nu=$1/5 is a FQHS \cite{Jiang.PRL.1990, Goldman.PRL.1993, Sajoto.PRL.1993}.

\begin{figure*}[t]
  \begin{center}
    \psfig{file=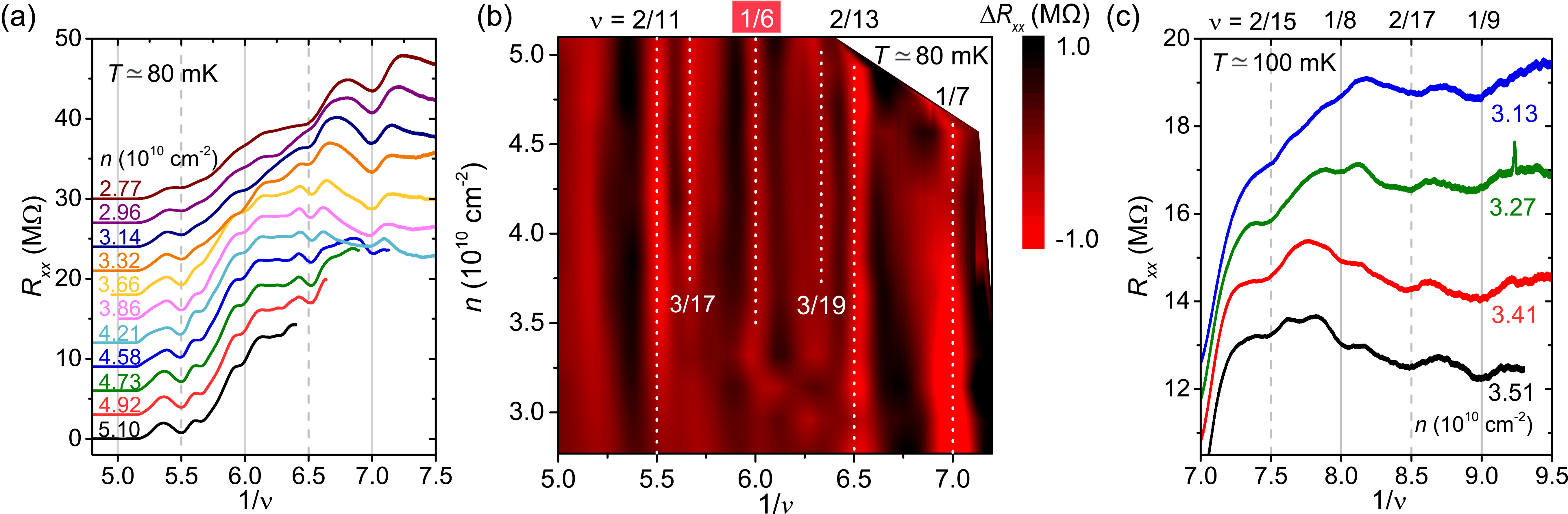, width=1 \textwidth}
  \end{center}
  \caption{\label{tilt} 
   {\bf Density dependence.} (a) $R_{xx}$ vs $1/\nu$ traces measured at $T\simeq$80 mK for different densities $n$. Each trace is vertically shifted by 3 M$\Omega$ for clarity. $n$ is tuned by symmetrically gating the 2DES from both the top and bottom. (b) Color-scale plot of $\Delta R_{xx}$ as a function of 1/$\nu$ and $n$. Several fillings are marked by white dotted lines. (c) $R_{xx}$ vs 1/$\nu$ traces measured at $T\simeq$100 mK and at very small $\nu$. $R_{xx}$ minima are observed at $\nu=$1/9, 2/17, 1/8, and 2/15.
   }
  \label{fig:tilt}
\end{figure*}

We measured a second sample from the same wafer \cite{Footnote.gate}. Figure 2(a) shows the $R_{xx}$ vs 1/$\nu$ traces measured at $T\simeq$80 mK with $n$ ranging from 2.77 to 5.10, while maintaining symmetric charge distribution. We observe a clear inflection point at $\nu=$1/6. It becomes weaker at lower densities \cite{Footnote.weaker features}. In addition, well-defined $R_{xx}$ minima are observed at $\nu=$2/11 in the whole range of $n$, and at $\nu=$1/7 in the lower density traces where we could reach $\nu=$1/7 with our magnet. Figure 2(b) displays a color-scale plot of $\Delta R_{xx}$ as a function of 1/$\nu$ and $n$. Distinct $\Delta R_{xx}$ minima are observed at $\nu=$2/11, 3/17, 3/19, 1/6, 2/13, and 1/7, with their 1/$\nu$ positions remaining consistent across different $n$, indicating that the signatures of the even-denominator FQHS at $\nu=$1/6, as well as the high-order Jain-sequence FQHSs at $\nu=p/(6p\pm1)$, are intrinsic to our ultrahigh-quality 2DES.

Figure 2(c) data show the fate of the FQHSs at yet smaller $\nu$. Here we observe $R_{xx}$ minima at odd-denominator $\nu=$1/9, 2/17, and 2/15. Hints of developing FQHSs were previously reported at $\nu=$1/9 by optical and transport measurements \cite{Buhmann.PRL.1990, Pan.PRL.2002}. Recent calculations also suggest that the ground state at $\nu=$1/9 is likely a FQHS \cite{Zuo.PRB.2020}. Our data revealing $R_{xx}$ minima at fixed fillings over a range of densities provide strong evidence for the existence of FQHSs at $\nu=p/(8p\pm1)$, namely at $\nu=$1/9, 2/17, and 2/15 \cite{Footnote.saturation}. Furthermore, the traces also exhibit an $R_{xx}$ minimum (or an inflection point) at the \textit{even-denominator} filling $\nu=$1/8. The qualitative resemblance of the data near $\nu=$1/8 in Fig. 2(c) to Fig. 2(a) data suggests that the physics for $^6$CFs for 1/7$<\nu<$1/5 can be extended to $^8$CFs for 1/9$<\nu<$1/7.

\begin{figure}[t]
  \begin{center}
    \psfig{file=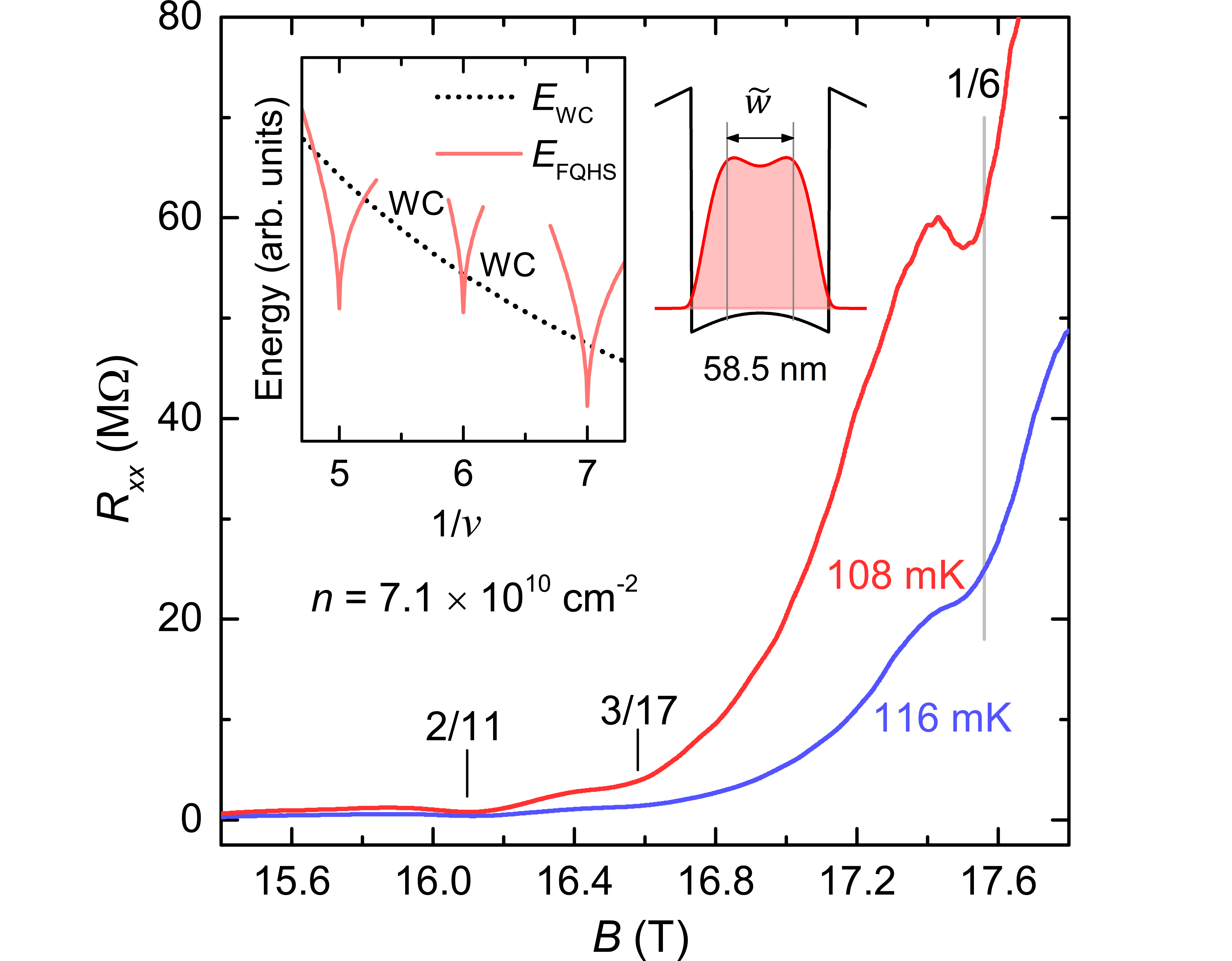, width=0.5\textwidth}
  \end{center}
  \caption{\label{other}
   {\bf Data for a different sample.} $R_{xx}$ vs $B$ traces for a sample with a density of 7.1$\times$10$^{10}$ cm$^{-2}$. Left inset: Schematic diagram showing the energies of WC and FQHSs vs $1/\nu$, indicating the possibility of a downward cusp in energy at $1/\nu=6$, similar to cusps at $1/\nu=5$ and 7. Right inset: Self-consistent charge distribution (red) and potential (black) for the 2DES confined to a 58.5-nm-wide QW; $\widetilde{w}$ denotes the electron layer thickness (see text).
   }
  \label{fig:other}
\end{figure}

The signatures of the $\nu=$1/6 FQHS are not specific to one wafer. In Fig. 3 we present data for another ultrahigh-quality GaAs 2DES from a different wafer with a higher density of 7.1 and a narrower QW width of 58.5 nm. We find a clear, sharp $R_{xx}$ minimum at $\nu=$1/6 at 108 mK \cite{Footnote.minimum}, and an inflection point at a slightly higher temperature of 116 mK.

\begin{table*} [t]
  \caption{\bf Sample parameters}
  \begin{ruledtabular}
    \begin{tabular}{ccccccc}
      Sample & \makecell{density \\ (10$^{10}$ cm$^{-2}$)} & \makecell{mobility \\ (10$^6$ cm$^2$/Vs)} & \makecell{QW width \\ (nm)} & \makecell{$\widetilde{w}/l_B$ \\ (at $\nu=$1/6)} & \makecell{$\Delta_{\text{SAS}}/E_{\text{Coul}}$ \\ (at $\nu=$1/6)} & \makecell{strength of $\nu=$1/6 \\ FQHS feature} \\
      \colrule
      A & 4.4 & 22 & 70 & 4.3 & 0.13 & Strong\\
      B & 7.1 & 25 & 58.5 & 4.6 & 0.15 & Strong\\
      C & 4.3 & 16 & 50 & 2.7 & 0.33 & Weak\\
      D & 4.5 & 17 & 40 & 2.2 & 0.52 & Weak\\
      E & 4.3 & 10 & 30 & 1.7 & 0.93 & Absent\\
    \end{tabular}
  \end{ruledtabular}
\end{table*}

One potential competitive ground state at $\nu=$1/6 is a metallic Fermi sea of $^6$CFs at zero effective magnetic field. However, several observations suggest that the $R_{xx}$ minimum we observe at $\nu=$1/6 is not indicative of a $^6$CF Fermi sea: (i) At $\nu=$1/2, where a Fermi sea of $^2$CFs is well established, typically a smooth and broad minimum in $R_{xx}$ is observed; see, e.g., Figs. S4 and S5 in SM \cite{SM}. In contrast, the $R_{xx}$ minimum at $\nu=$1/6 is sharp; see Figs. \ref{fig:main} and \ref{fig:other}, and also the sharp peak in $d^2R_{xx}/dB^2$ [Fig. S2(b)]. The sharp $R_{xx}$ minimum observed in the WC regime at $\nu=1/6$ indicates that the $\nu=1/6$ state is flanked by WC states. This strongly favors the interpretation of FQHS over $^6$CF Fermi sea: The energies of FQHSs show downward cusps as a function of $\nu$, and can dip below the WC energy (see Fig. \ref{fig:other} left inset, and also see Fig. 9 in Ref. \cite{Shayegan.review.2006}) so that the WC could be reentrant on the flanks of a FQHS \cite{Archer.PRL.2013, Jiang.PRL.1990, Zuo.PRB.2020}. On the other hand, the energies of both CF Fermi sea and WC states are smooth functions of $\nu$, and thus one would expect only a single transition between these two states. (ii) The $R_{xx}$ minimum at $\nu=$1/6 and its temperature dependence closely resemble those at Jain-sequence, odd-denominator fillings, such as $\nu=$1/7 (see Fig. 1 and also Fig. S6 in SM \cite{SM, Footnote.activation}), challenging the metallic $^6$CF Fermi sea interpretation. (iii) A broad $R_{xx}$ minimum is not a universal feature for a CF Fermi sea, and is indeed generally absent at $\nu=$1/4 when there is a competing insulating phase or rising background resistance \cite{Pan.PRB.2000}; also see SM \cite{SM}. It is also worth noting that the observation of Jain-sequence FQHSs does not necessarily indicate the presence of a CF Fermi sea. Indeed, a FQHS at $\nu=1/2$ flanked by numerous Jain-sequence FQHSs is well established in 2DESs confined to wide GaAs QWs \cite{Suen.PRL.1992, Suen.PRL.1994, Shabani.PRL.2009, Shabani.PRB.2013, Mueed.PRL.2015, Mueed.PRL.2016, Singh.NatPhys.2024}. We emphasize that, while we believe our data favor the presence of a developing FQHS at $\nu=$1/6, taking the $R_{xx}$ minimum as evidence for a $^6$CF Fermi sea at 1/6 is also novel as there has been no experimental or theoretical evidence for such a phase deep in the low-filling, insulating, WC regime.

The emergence of even-denominator FQHSs at $\nu=$1/6 and 1/8 is unexpected. These states have not been reported in any 2D carrier system, or predicted by existing theories. One might wonder if these new enigmatic states share a common origin with two other even-denominator FQHSs, namely those at $\nu=$1/2 and 1/4, observed in the lowest LL in wide QWs under special circumstances \cite{Suen.PRL.1992, Suen.PRL.1994, Luhman.PRL.2008, Shabani.PRL.2009, Shabani.PRB.2013, Mueed.PRL.2015, Mueed.PRL.2016, Singh.NatPhys.2024}. The origin of these states has been a subject of debate \cite{Peterson.PRB.2010}; however, recent experiments \cite{Mueed.PRL.2015, Mueed.PRL.2016, Singh.NatPhys.2024} and theories \cite{Zhu.PRB.2016, Faugno.PRL.2019, Sharma.PRB.2024} indicate they are likely single-component, non-Abelian states arising from a pairing instability of CFs. To examine this possibility and investigate the role of QW width on the 1/6 FQHS, we studied several samples with similar densities $n=$4.4, 4.3, 4.5 and 4.3, but different QW widths of 70, 50, 40, and 30 nm, respectively. Below we summarize our findings; for details, see Sec. IV in SM \cite{SM}. 

Table I contains the relevant parameters for our samples. One parameter is the effective electron layer thickness ($\widetilde{w}$), normalized to the magnetic length $l_B=\sqrt{\hbar/eB}$;  $\widetilde{w}$ is typically defined as twice the standard deviation of the calculated charge distribution from its center; see Fig. \ref{fig:other} inset for an example. Another parameter is the symmetric-to-antisymmetric subband separation ($\Delta_{\text{SAS}}$) normalized to the Coulomb energy, $E_{\text{Coul}}=e^2/(4\pi\epsilon l_B)$. For 2DESs confined to QWs, in the extreme quantum limit ($\nu<$1), although only one LL that originates from the symmetric subband is partially occupied, the proximity of the antisymmetric subband can potentially modify the electron-electron interaction in the lowest LL when $\Delta_{\text{SAS}}/E_{\text{Coul}}\ll$1. As seen in Table I, we observe the strongest $\nu=$1/6 FQHS in samples A and B where $\widetilde{w}/l_B$ is large ($\gtrsim$4) and $\Delta_{\text{SAS}}/E_{\text{Coul}}$ is small ($\lesssim$0.2), suggesting that the $\nu=$1/6 FQHS is stabilized by the large layer thickness and proximity of the antisymmetric subband. 

While the 1/2 and 1/4 FQHSs are also seen at large $\widetilde{w}/l_B$ and reasonably small $\Delta_{\text{SAS}}/E_{\text{Coul}}$, the 1/6 data pose several notable quantitative and qualitative differences. First, the parameters for the samples where we observe the strong $\nu=$1/6 minima are different compared to those where the 1/2 and 1/4 FQHSs are seen. This is visually exhibited in “phase diagrams” for the stability of the $\nu=$1/2 FQHS in wide GaAs QWs shown in Fig. S9 in SM \cite{SM}. Second, the 1/2 and 1/4 FQHSs are observed only in samples with \textit{bilayer} charge distributions \cite{Suen.PRL.1992, Suen.PRL.1994, Luhman.PRL.2008, Shabani.PRL.2009, Shabani.PRB.2013, Mueed.PRL.2015, Mueed.PRL.2016, Singh.NatPhys.2024}, but in our samples the charge distribution is single-layer, albeit thick [see Fig. 1(c) and Fig. 3 inset]. Third, the 1/2 and 1/4 FQHSs in wide QWs become weaker and disappear when the charge distribution is made asymmetric \cite{Suen.PRL.1994, Manoharan.PRL.1996, Shabani.PRL.2009}. In contrast, the developing 1/6 FQHS in our sample is robust against asymmetry (see Figs. S10 and S11 \cite{SM}). Fourth, at a fixed QW width, by increasing $n$, or equivalently $\widetilde{w}/l_B$, the ground state at $\nu=$1/2 in wide QWs starts as a compressible Fermi sea, transitions to an incompressible FQHS, and eventually turns into an insulating phase \cite{Suen.PRL.1994, Manoharan.PRL.1996, Shabani.PRB.2013}. At $\nu=$1/6, however, the 2DES is deep in the insulating regime, and a developing FQHS emerges when $\widetilde{w}/l_B$ becomes large.

It is worth reiterating that, while the origin of the 1/2 FQHS in wide QWs is not entirely clear, recent experiments \cite{Mueed.PRL.2015, Mueed.PRL.2016, Singh.NatPhys.2024} and theories \cite{Zhu.PRB.2016, Sharma.PRB.2024} point to a single-component, non-Abelian, Pfaffian-like state stabilized by a $p$-wave pairing of $^2$CFs. Similarly, the 1/4 FQHS has been interpreted as a single-component, non-Abelian FQHS stabilized by $f$-wave pairing of $^4$CFs \cite{Faugno.PRL.2019, Sharma.PRB.2024}. Despite the differences enumerated in the previous paragraph between the parameters of our samples and those which show the 1/2 and 1/4 FQHSs, we suggest that the developing 1/6 FQHS we report here is also single-component and non-Abelian, and is stabilized by a pairing of $^6$CFs [see Fig. 1(d)]. An alternative, two-component ground state, such as the Halperin-Laughlin $\Psi_{993}$ state \cite{MacDonald.SurfSci.1990}, is extremely unlikely to be the ground state, given the single-layer-like charge distribution in our samples and the persistence of the 1/6 $R_{xx}$ minimum as we make the charge distribution significantly asymmetric. Consistent with our conjecture that the 1/6 FQHS we are observing has a one-component origin, preliminary theoretical calculations indeed suggest that this state likely emerges from an $f$-wave pairing of $^6$CFs and hosts non-Abelian quasiparticles \cite{Sharma.Balram.Jain.unpublished}. We hope, of course, that our results will stimulate future theoretical calculations to shed more light on the origin of the FQHSs at $\nu=$1/6 (and 1/8).

\begin{acknowledgements}
\textit{Acknowledgments} --- We acknowledge support by the National Science Foundation (NSF) Grants No. DMR 2104771 and No. ECCS 1906253) for measurements, the U.S. Department of Energy (DOE) Basic Energy Sciences Grant No. DEFG02-00-ER45841) for sample characterization, the Eric and Wendy Schmidt Transformative Technology Fund, and the Gordon and Betty Moore Foundation’s EPiQS Initiative (Grant No. GBMF9615.01 to L.N.P.) for sample fabrication. Our measurements were partly performed at the National High Magnetic Field Laboratory (NHMFL), which is supported by the NSF Cooperative Agreement No. DMR 2128556, by the State of Florida, and by the DOE. This research was funded in part by QuantEmX grant from Institute for Complex Adaptive Matter and the Gordon and Betty Moore Foundation through Grant GBMF9616 to C. W., A. G., P. T. M., and S. K. S. We thank A. Bangura, G. Jones, R. Nowell and T. Murphy at NHMFL for technical assistance, and Ajit C. Balram and Jainendra K. Jain for illuminating discussions.
\end{acknowledgements}



\section*{Supplementary material}

# Supplemental Material to "Developing fractional quantum Hall states at even-denominator fillings 1/6 and 1/8"

Chengyu Wang,[1] P. T. Madathil,[1] S. K. Singh,[1] A. Gupta,[1] Y. J. Chung,[1] L. N. Pfeiffer,[1] K. W. Baldwin,[1] and M. Shayegan[1]

[1]*Department of Electrical and Computer Engineering, Princeton University, Princeton, New Jersey 08544, USA*

(Dated: November 13, 2024)

# I. TWO METHODS OF EXTRACTING FRACTIONAL QUANTUM HALL STATE FEATURES FROM HIGHLY INSULATING LONGITUDINAL RESISTANCE

In order to better visualize the fractional quantum Hall state (FQHS) features in $R_{xx}$ amidst the dominant insulating behavior caused by the competing Wigner crystal (WC) states, we extract the oscillatory component ($\Delta R_{xx}$) of the $R_{xx}$ data. Here $\Delta R_{xx}$ is defined as the resistance after subtracting the smooth background. Figures S1(a-h) display the raw $R_{xx}$ traces (black) and their corresponding smooth backgrounds (red) obtained using second-order Savitzky-Golay smoothing with a window size of 1.5 T. This window size captures the main trend of the slowly varying part of $R_{xx}$ without introducing artificial features. To ensure consistency across different densities, the window sizes we use for obtaining $\Delta R_{xx}$ in Fig. 2(b) of the main text are scaled according to the density. Figure S1(i) shows a color-scale plot of $\Delta R_{xx}$ as a function of $B$ and $T$. In Fig. S1(j) we show the same set of data by plotting $\Delta R_{xx}$ vs $B$ traces at different temperatures. Clear minima are observed at $\nu = 2/11$, $3/17$, $4/23$, $1/6$, $4/25$, $3/19$, $2/13$, and $1/7$, suggesting the emergence of FQHSs at these fillings. The data reveal that the odd-denominator FQHS features become less robust as the filling factor approaches $\nu = 1/6$, consistent with what is generally observed for standard Jain-sequence FQHSs. Furthermore, the $\Delta R_{xx}$ minimum at $\nu = 1/6$ is sharp, and shows a strong temperature dependence. This is very similar to the $\Delta R_{xx}$ minima at odd-denominator fillings on the flanks of $\nu = 1/6$, but very different compared to the smooth, broad $R_{xx}$ minimum at $\nu = 1/2$ which is rather insensitive to small temperature changes.

Another effective method to extract FQHS features from highly insulating $R_{xx}$ data is to calculate the second derivative, $d^2R_{xx}/dB^2$. A developing FQHS manifests itself as a sharp, maximal curvature in $R_{xx}$, and therefore exhibits a sharp peak in $d^2R_{xx}/dB^2$. In Fig. S2, we compare the color-scale plot of $d^2R_{xx}/dB^2$ as a function of $B$ and $T$ [panel (b)] to data in Fig. S1(i) [replotted as panel (a) of Fig. S2]. Both methods reveal sharp FQHS features at $\nu = 1/6$ and odd-denominator, Jain-sequence fillings $\nu = p/(6p \pm 1)$. As another example of the similarity between $\Delta R_{xx}$ and $d^2R_{xx}/dB^2$, in Fig. S3 we show color-scale plots of $\Delta R_{xx}$ [panel (a)] and $d^2R_{xx}/dB^2$ [panel (b)] as a function of $1/\nu$ and $n$.

## II. MAGNETO-TRANSPORT DATA NEAR LANDAU LEVEL FILLINGS 1/2 AND 1/4

Figure S4(b) shows $R_{xx}$ vs $B$ traces measured with $I = 20$ nA at $T \simeq 55$ mK for sample A. The trace shows a broad, smooth minimum at $\nu = 1/2$, flanked by the Jain-sequence FQHSs of two-flux composite fermions ($^2$CFs) at $\nu = p/(2p \pm 1)$; $p = 1, 2, 3, ...$, e.g., at $\nu = 1/3, 2/5, 3/7, ..., 9/19$, and $\nu = 2/3, 3/5, 4/7, ..., 9/17$. The trace at higher fields (blue) is rather featureless near $\nu = 1/4$, with Jain-sequence FQHSs of four-flux CFs ($^4$CFs) observed on the flanks at $\nu = p/(4p \pm 1)$. Figure S4(a) presents the second derivative of the data shown in Fig. S4(b). Sharp peaks in $d^2R_{xx}/dB^2$ are observed flanking $\nu = 1/2$ and $\nu = 1/4$, e.g., at $\nu = 9/17, 9/19, 6/23$, and $5/21$, consistent with developing Jain-sequence FQHSs of $^2$CFs and $^4$CFs. At $\nu = 1/2$ and $\nu = 1/4$, $d^2R_{xx}/dB^2$ is featureless and remains close to zero over a range of $B$, consistent with compressible CF Fermi sea ground states at $\nu = 1/2$ and $\nu = 1/4$ [1–4]. This is in stark contrast to the sharp peaks observed at odd-denominator Jain-sequence fillings and the even-denominator $\nu = 1/6$ [see Figs. S2(b) and S3(b)].

In Fig. S5, we present $R_{xx}$ data for sample B. Qualitatively similar behaviors near $\nu = 1/2$ and 1/4 are also observed in sample B. Compared to sample A, higher-order Jain-sequence FQHSs in sample B are observed on the flanks of $\nu = 1/2$ up to $\nu = 12/23$ and 12/25, thanks to the lower sample temperature ($\simeq 40$ mK) and higher 2D electron density ($7.1 \times 10^{10}$ cm$^{-2}$). We note that sample B exhibits a weak, broad maximum at $\nu = 1/2$. The origin of this maximum is unclear. No $R_{xx}$ minimum is observed at $\nu = 1/4$. $d^2R_{xx}/dB^2$ is featureless and remains close to zero near $\nu = 1/2$ and 1/4.

## III. TEMPERATURE DEPENDENCE AND ACTIVATION ENERGY NEAR LANDAU LEVEL FILLING 1/6

Figure S6(a) shows the high-field $R_{xx}$ vs $B$ traces taken at different temperatures ranging from 80 mK to 142 mK. In Fig. S6(b), we present activation energy ($E_A$) data, deduced from the relation $R_{xx} \propto e^{E_A/2kT}$, as a function of $B$. Figure S6(c) shows $R_{xx}$ vs $1/T$ Arrhenius plots and the linear fits whose slopes give the value of $E_A$ at a given $B$. The magnitude of $E_A$ is associated with the defect formation energy of WC. Numerical calculations indicate

that, at high $B$, a WC formed by $^4$CFs has lower energy than the conventional Hartree-Fock electron crystal [5], and that the lowest-energy *intrinsic* defect in this $^4$CF WC is a hyper-correlated quantum bubble defect [6]. Recent measurements [7] on ultrahigh-quality two-dimensional electron systems (2DESs) similar to our sample, indeed exhibit $E_A$ values which are in agreement with what is expected for an ideal $^4$CF WC with no disorder, consistent with the fact that the WC domains in these samples are estimated to contain $\simeq$ 1000 electrons [8].

Note that Fig. S6(b) data also reveal clear minima at $\nu =$ 1/7, 2/11, and 1/6, where signatures of developing FQHSs are observed in the $R_{xx}$ vs $B$ data in Figs. S6(a). Similar minima in $E_A$ were reported in modest-quality GaAs 2DESs at $\nu = 1/5$ [9] and very recently in ultrahigh-quality GaAs 2DESs at $\nu = 1/7$ [10], and were interpreted as precursors to developing FQHSs. The observation of an $E_A$ minimum at $\nu = 1/6$ in Fig. S6(b) therefore supports the notion that the ground state at this filling is a FQHS.

## IV. DATA FOR SAMPLES WITH DIFFERENT GaAs QUANTUM WELL WIDTHS

To investigate the role of quantum well (QW) width on the 1/6 FQHS, we studied several samples with similar densities $n =$ 4.4, 4.3, 4.5 and 4.3 (in units of $10^{10}$ cm$^{-2}$), but different QW widths of 70, 50, 40, and 30 nm, respectively; see Table I in the main text for sample parameters. In all the samples, we observe clear $R_{xx}$ minima or inflection points at $\nu =$ 1/7, 2/13, and 2/11; see Figs. S6(a) and S7. They become more pronounced at lower temperatures, signaling the emergence of developing Jain-sequence FQHSs of $^6$CFs. These features attest to the exceptionally high quality of all the samples.

At $\nu = 1/6$, the FQHS feature becomes weaker in narrower QWs: In contrast to the clear $R_{xx}$ minimum we observe in the 70-nm-wide sample, the 50 and 40-nm-wide samples exhibit a weak inflection point in $R_{xx}$ [also evidenced by the sharp peak in $d^2R_{xx}/dB^2$ centered at $\nu = 1/6$; see Figs. S7(b, c)], while the 30-nm-wide sample is featureless near $\nu = 1/6$; see Fig. S7(d). In Fig. S7, we do not show data at lower $T$ because the out-of-phase (capacitive) signal becomes comparable or even larger than the in-phase (resistive) signal and $R_{xx}$ cannot be measured reliably near $\nu = 1/6$.

In Fig. S8, we present low-$B$ traces for the three samples shown in Figs. S7(b-d). Qualitatively similar behaviors are observed in these samples: A smooth, shallow $R_{xx}$ minimum

4

is seen at $\nu = 1/2$, while no $R_{xx}$ minimum is observed at $\nu = 1/4$. These observations are consistent with what we see in Figs. S4 and S5 for the 70-nm-wide and the 58.5-nm-wide samples (samples A and B).

## V. COMPARISON WITH PHASE DIAGRAMS FOR 1/2 FQHS IN WIDE GaAs QUANTUM WELLS

In order to contrast the relevant parameters of our samples with those which exhibit the $\nu = 1/2$ and $1/4$ FQHSs, we present Fig. S9. This figure contains "phase diagrams" for the stability of the $\nu = 1/2$ FQHS in wide GaAs QWs [11]. Figure S9(a) shows a plot of QW well width ($w$) vs density $n$, with the yellow-shaded area marking the parameter range where the $\nu = 1/2$ FQHS is stable, and the cyan and blue circles denoting where $\nu = 1/4$ FQHS has been reported [12, 13]. We have also included two data points (red stars) for the parameters of our samples A and B which exhibit clear minima at $\nu = 1/6$. In Fig. S9(b), we show the same phase diagrams and data, but now the axes are $\widetilde{w}/l_B$ and $\Delta_{SAS}/E_{Coul}$.

## VI. IMBALANCE DATA

In Fig. S10, we present $R_{xx}$ vs $1/\nu$ traces for the 70-nm-wide QW at a fixed density of $4.6 \times 10^{10}$ cm$^{-2}$ with imbalanced charge distributions. (Front and back gates with opposite polarities were used to achieve these imbalanced charge distributions.) We observe $R_{xx}$ minima at $\nu = 2/11$, $2/13$, and $1/7$. A clear inflection point is observed at $\nu = 1/6$ in all traces. Figure S11 shows $\Delta R_{xx}$ vs $1/\nu$ traces, revealing the evolution of FQHS features with imbalanced charge distributions more clearly. Sharp $\Delta R_{xx}$ minima are observed at $\nu = 2/11$, $3/17$, $2/13$, $1/7$, and $1/6$ in all traces. The FQHS features at odd-denominator fillings $\nu = 2/11$, $3/17$, $2/13$, and $1/7$ are rather insensitive to charge imbalance, consistent with what is expected for one-component, Jain-sequence FQHSs in a single-layer system. The FQHS feature at $\nu = 1/6$ is also robust against large charge asymmetry ($\Delta n/n \sim 50\%$), suggesting that the $\nu = 1/6$ FQHS we observe also has a one-component origin.

## VII. HALL DATA

In Fig. S12, we present Hall slope ($dR_H/dB$) and Hall resistance ($R_H$) for sample A. We observe flat quantized Hall plateaus at integer fillings $\nu = 1, 2, ...$ and fractional fillings, e.g., at $\nu = 1/3$ and $1/5$. Weaker Jain-sequence FQHSs that are not fully developed manifest themselves as developing plateaus in $R_H$ and sharp minima in $dR_H/dB$, e.g., at $\nu = 6/11$ and $6/13$. Near $\nu = 1/2$ and $1/4$, $R_H$ is linear and $dR_H/dB$ is featureless, consistent with metallic Fermi seas of $^2$CF and $^4$CF. It is worth noting that for $\nu < 1/5$ and $1/5 < \nu < 2/9$, $R_H$ significantly deviates from the classical Hall line and $dR_H/dB$ exhibits a complex behavior. This is because of the severe $R_{xx}$ mixing into the $R_H$ measurement when the 2DES enters a highly insulating regime where WC states prevail [14, 15], and because of the loss of Ohmic contact to the 2DES when $R_{xx}$ becomes immeasurably large.

To reduce the mixing of $R_{xx}$ and more accurately measure $R_H$ at extremely small $\nu \sim 1/6$, we perform Hall measurements at elevated temperatures, where the ratio $R_{xx}/R_H$ is smaller. Figure S13 shows $R_H$ and $dR_H/dB$ traces taken at two different temperatures. At $T \simeq 150$ mK, the Hall plateaus and the corresponding $dR_H/dB$ minima at low $B$ (e.g., at $\nu = 1$, $1/3$, and $1/5$) are much weaker compared to the low temperature data shown in Fig. S12. Here, $R_H$ begins to show mixing from $R_{xx}$ and significantly deviates from the classical Hall line near $\nu \sim 1/6$. However, no features are observed in the range $1/5 > \nu > 1/6$. At $T \simeq 177$ mK, $R_H$ follows the classical Hall line down to a lower filling ($\nu \sim 1/7$). Again, no features are observed in the range $1/5 > \nu > 1/7$.

It is important to note that Jain-sequence FQHSs of $^6$CFs are much more fragile compared to those of $^2$CFs and $^4$CFs. Theoretical studies report gap energies of approximately 0.1, 0.03, and 0.007 (in units of the Coulomb energy, $e^2/4\pi\epsilon\epsilon_0 l_B$) for the $\nu = 1/3$, $1/5$, and $1/7$ FQHSs, respectively [16]. Consistent with this trend, in Fig. S13, the Hall plateau at $\nu = 1/5$ is indeed more fragile against increasing temperature than the $\nu = 1/3$ plateau. At even smaller fillings in the regime of $^6$CFs, the strength of the $R_{xx}$ minimum at $\nu = 1/6$ is comparable to those at $\nu = 2/13$ and $3/17$ [see Fig. 1 of main text]. They get weaker rapidly and eventually disappear when $T$ is increased from 80 mK to 142 mK [Fig. 1], in stark contrast to the $\nu = 1/5$ $R_{xx}$ minimum which survives up to $T > 300$ mK. Therefore, the absence of Hall plateaus at $\nu = 1/6$ and Jain-sequence fillings $\nu = 1/7$, $2/13$, $3/19$, and $2/11$, $3/17$ at elevated temperatures is not unexpected.

In Fig. S14, we present $R_H$ data measured at intermediate temperatures, where $R_{xx}$ minima or inflection points are observed at the even-denominator filling factor $\nu = 1/6$ and the odd-denominator filling factors $\nu = 1/7$, $2/13$, $3/17$, and $2/11$. At these temperatures, $R_H$ exhibits a plateau quantized at $5h/e^2$ and a pronounced minimum in $dR_H/dB$ developing at $\nu = 1/5$. At higher $B$, significant deviations of $R_H$ from the classical Hall resistance $B/ne$, caused by the mixing of $R_{xx}$, are clearly visible. The magnitudes of these deviations show sharp minima at $\nu = 1/7$, $2/13$, $3/17$, $2/11$, and $\nu = 1/6$, which qualitatively mirror the $R_{xx}$ data observed at similar temperatures. We also note that, as we increase the temperature, these sharp minima weaken and disappear before the mixing of $R_{xx}$ is eliminated.

To conclude, there are two factors that prevent us from measuring quantized Hall plateaus at extremely small $\nu$: (i) At low temperatures, Hall resistance cannot be measured reliably because of the severe $R_{xx}$ mixing effect caused by the dominant insulating phase. (ii) At higher temperatures, the fragile FQHSs at extremely small fillings (e.g., at $\nu = 1/7$, $2/11$ and $1/6$) diminish in strength and eventually vanish. Unfortunately, there is no temperature range within which we can reliably measure Hall resistance while the FQHSs at small $\nu$ still survive.

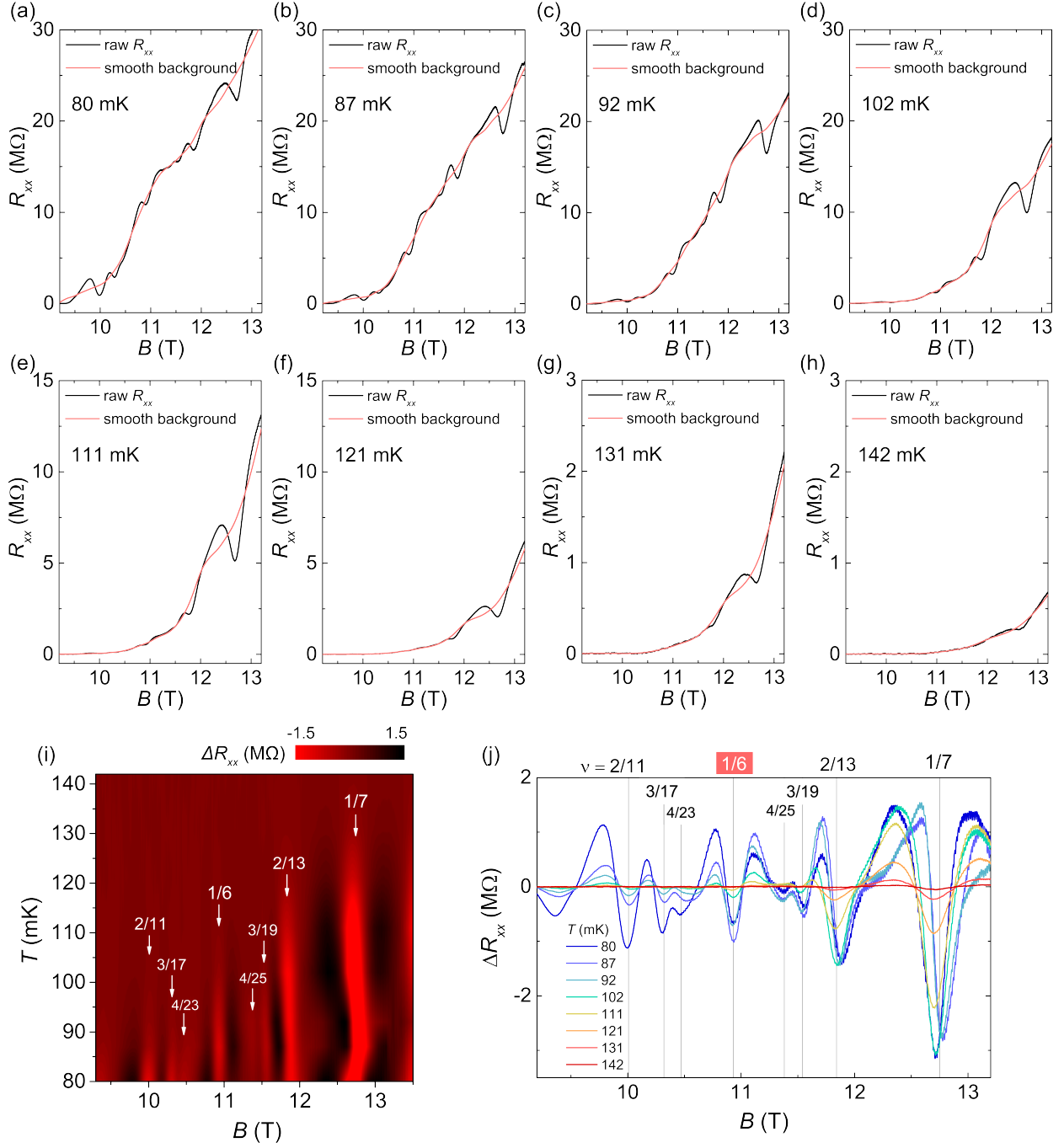

Fig. S1. **Extracting FQHS features from $R_{xx}$ with an increasing resistance background.** (a-h) $R_{xx}$ vs $B$ traces (black) and resistance background (red) obtained using Savitzky–Golay smoothing. (i) Color-scale plot of $\Delta R_{xx}$ as a function of $B$ and $T$, where $\Delta R_{xx}$ is the resistance after subtracting the increasing background. (j) $\Delta R_{xx}$ vs $B$ traces at different $T$.

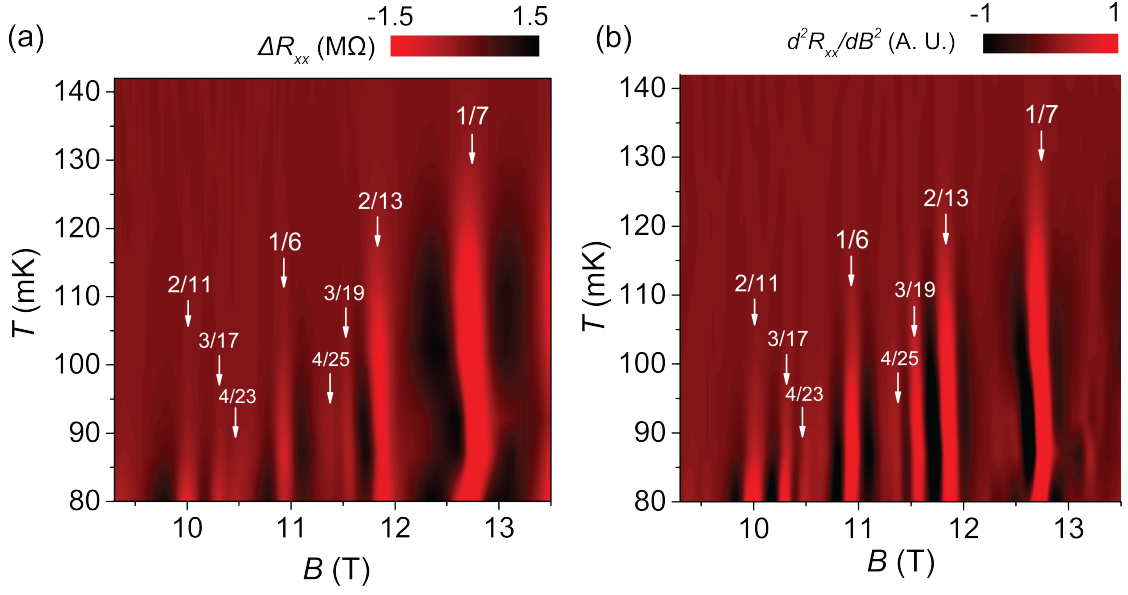

Fig. S2. **Comparison between $\Delta R_{xx}$ and $d^2 R_{xx}/dB^2$ data.** (a) Color-scale plot of $\Delta R_{xx}$ as a function of $B$ and $T$, adapted from Fig. S1(i). (b) Color-scale plot of $d^2 R_{xx}/dB^2$ as a function of $B$ and $T$.

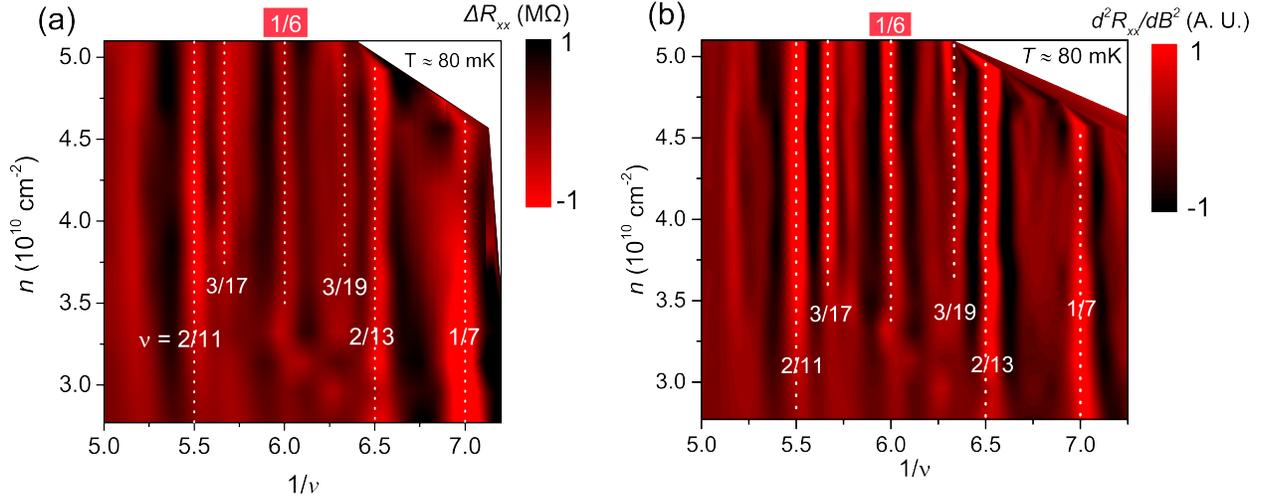

Fig. S3. **Comparison between $\Delta R_{xx}$ and $d^2 R_{xx}/dB^2$ data.** (a) Color-scale plot of $\Delta R_{xx}$ as a function of $1/\nu$ and $n$, adapted from Fig. 2(b) of the main text. (b) Color-scale plot of $d^2 R_{xx}/dB^2$ as a function of $1/\nu$ and $n$.

10

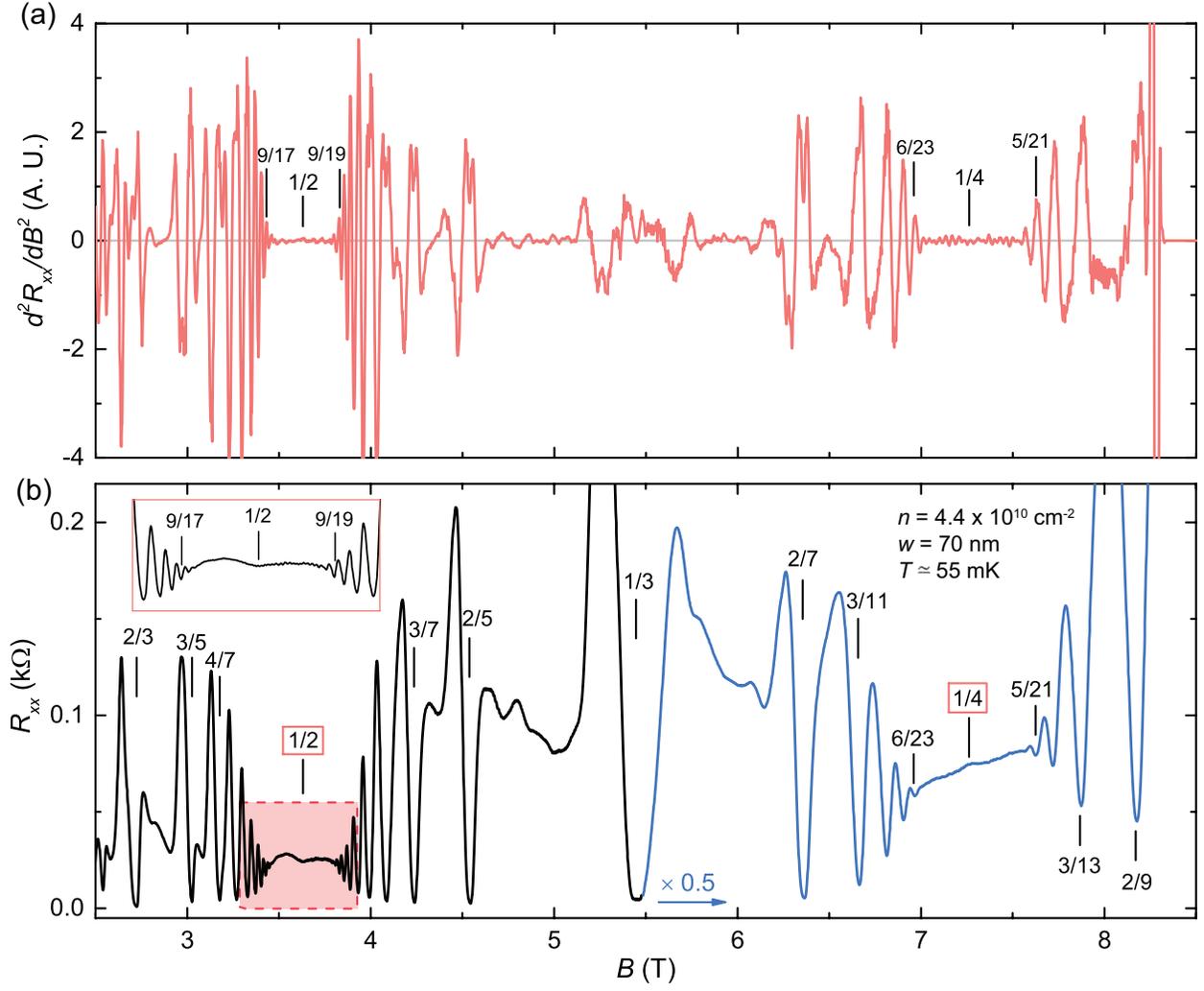

Fig. S4. **Magnetoresistance near $\nu = 1/2$ and $1/4$ for sample A.** (a) $d^2R_{xx}/dB^2$ and (b) $R_{xx}$ vs $B$ traces measured with $I = 20$ nA at $T \simeq 55$ mK. In (b), the magnitude of the trace for $B > 5.5$ T is multiplied by 0.5.

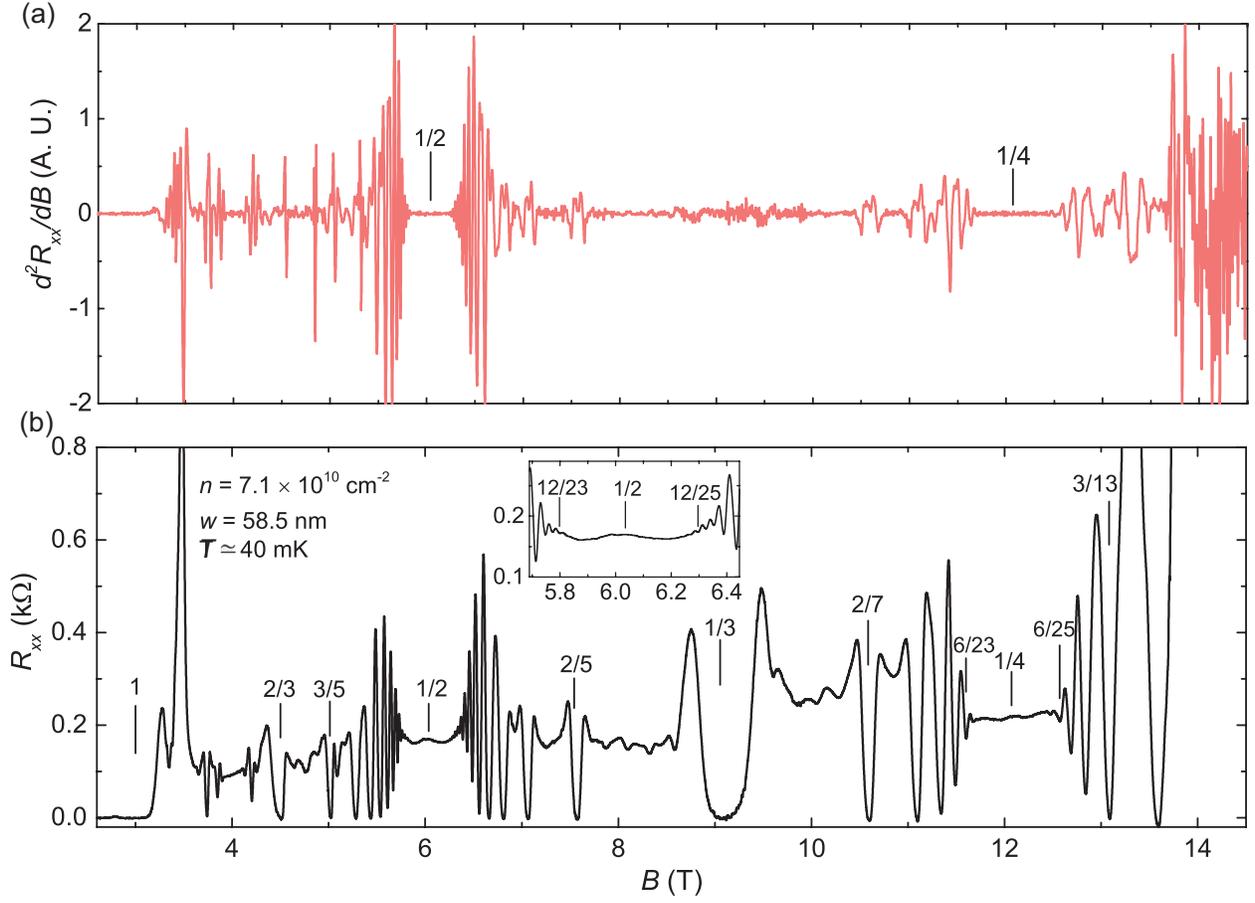

Fig. S5. **Magnetoresistance near $\nu = 1/2$ and $1/4$ for sample B.** (a) $d^2R_{xx}/dB^2$ and (b) $R_{xx}$ vs $B$ traces measured with $I = 20$ nA at $T \simeq 40$ mK.

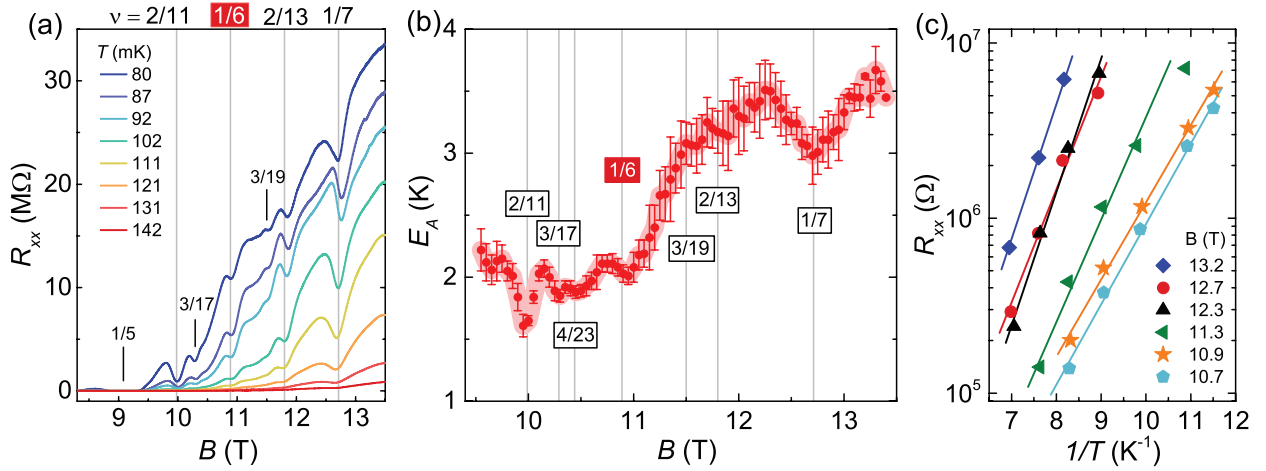

Fig. S6. **Temperature dependence and activation energy.** (a) $R_{xx}$ vs $B$ traces taken during up sweeps of $B$ at different temperatures. Several Landau level fillings are marked. (b) Activation energy $E_A$ deduced from the relation $R_{xx} \propto e^{E_A/2kT}$. (c) $R_{xx}$ vs $1/T$ Arrhenius plots at various $B$. The solid lines show linear fits whose slopes give the value of $E_A$.

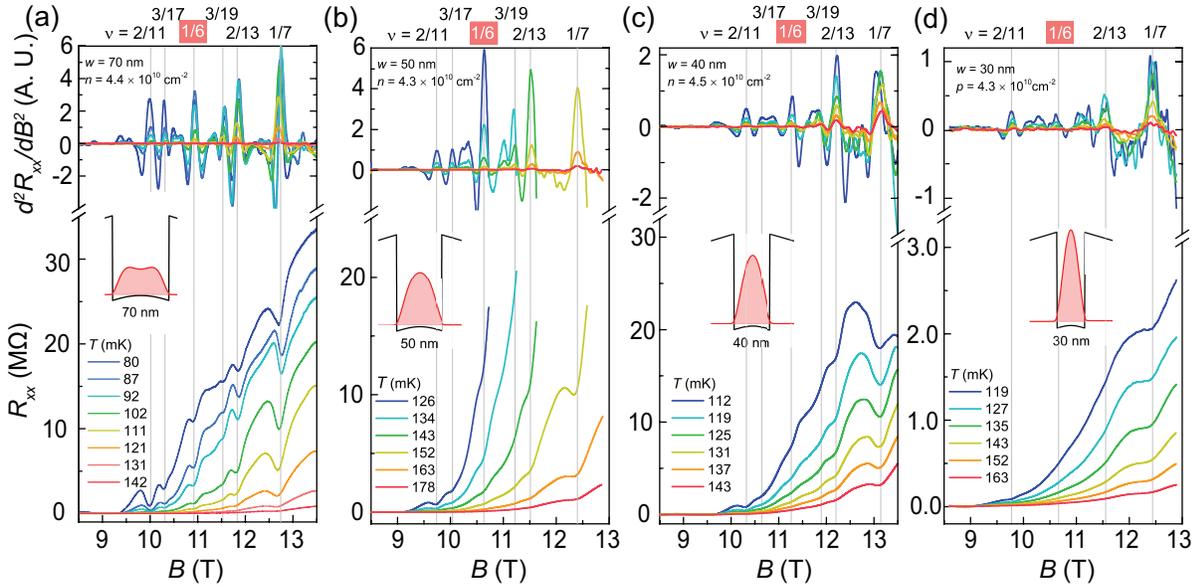

Fig. S7. **Data near $\nu = 1/6$ for samples with different GaAs QW widths.** $R_{xx}$ and $d^2R_{xx}/dB^2$ vs $B$ traces taken at different temperatures for: (a) $w = 70$ nm, $n = 4.4$; (b) $w = 50$ nm, $n = 4.3$; (c) $w = 40$ nm, $n = 4.5$; (d) $w = 30$ nm, $n = 4.3$. Insets: Self-consistent charge distributions (red) and potentials (black) for 2DESs confined to QWs with different widths.

13

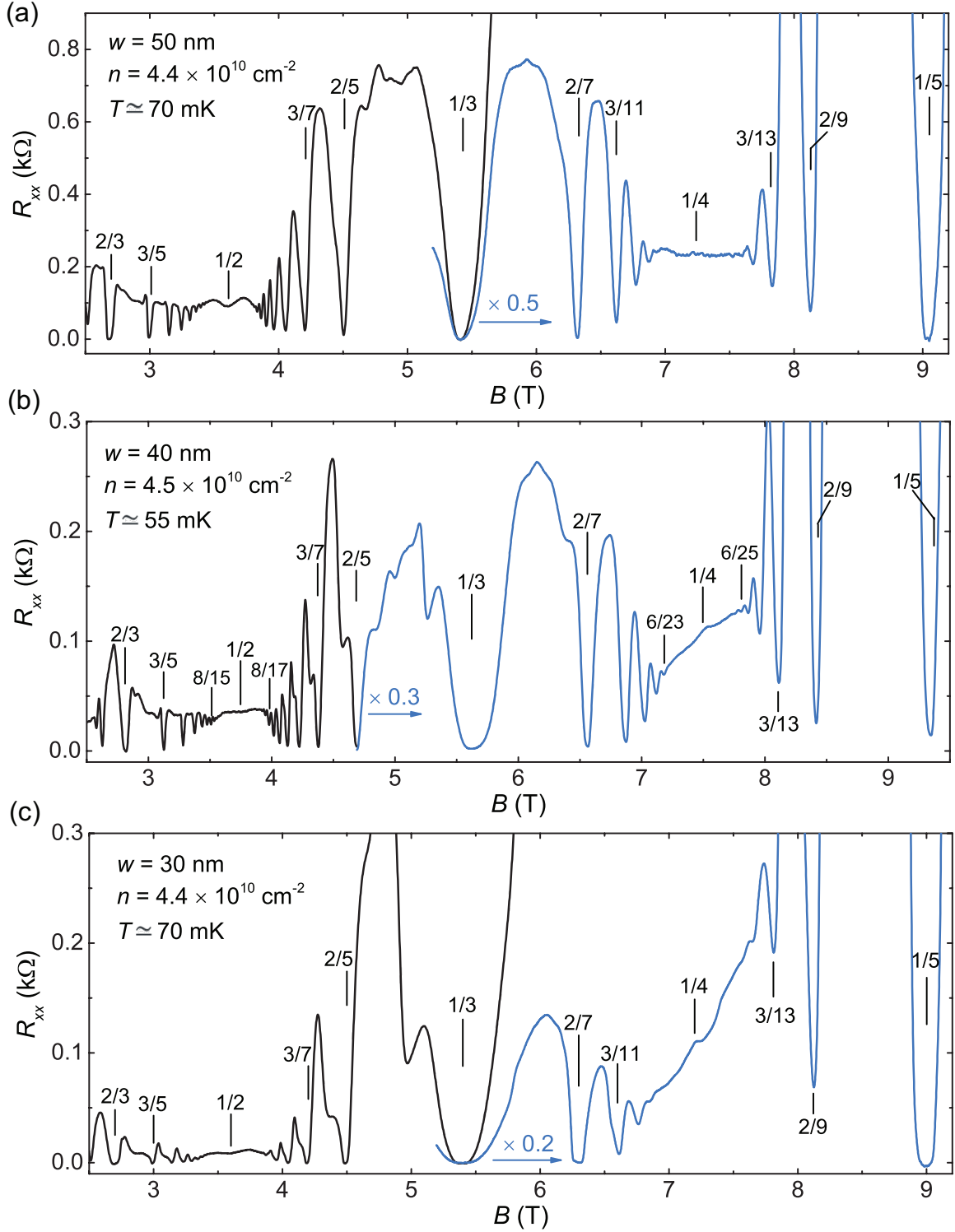

Fig. S8. **Data near $\nu = 1/2$ and $\nu = 1/4$ for samples with different GaAs QW widths.** $R_{xx}$ vs $B$ traces for three different samples: (a) $w = 50$ nm, $n = 4.3$; (b) $w = 40$ nm, $n = 4.5$; (c) $w = 30$ nm, $n = 4.3$.

14

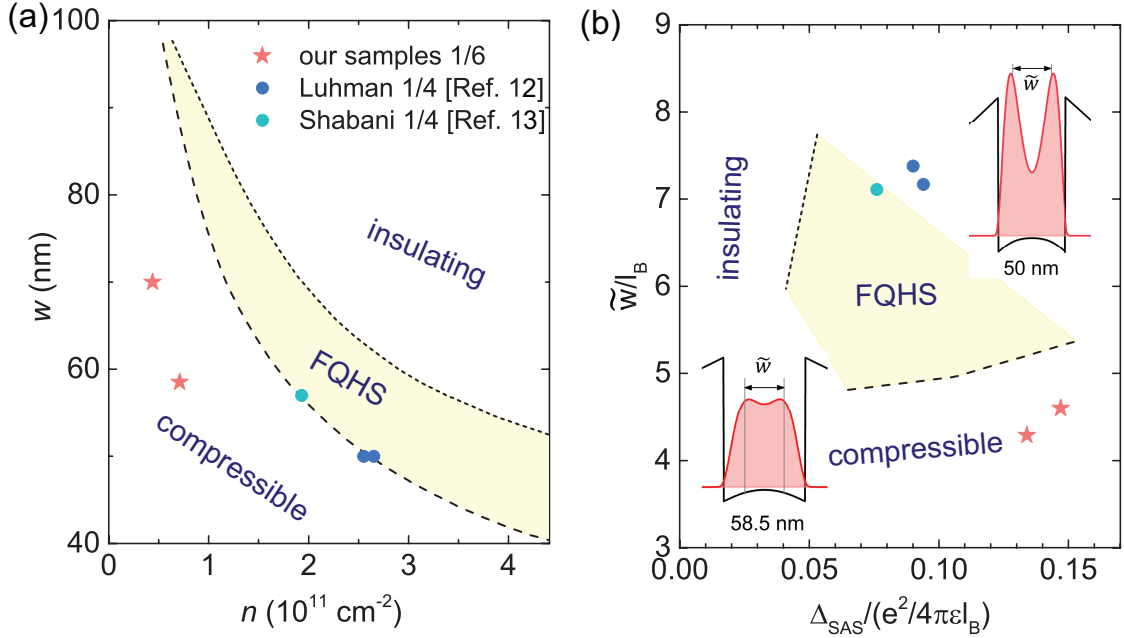

Fig. S9. **Comparison with phase diagrams for 1/2 FQHS in wide GaAs QWs.** (a) $w$ vs $n$, and (b) $\widetilde{w}/l_B$ vs $\Delta_{SAS}/(e^2/4\pi\epsilon l_B)$ phase diagrams showing the region, marked in yellow, in which the 1/2 FQHS is stabilized. As described in the main text, $\widetilde{w}$ is the effective electron layer thickness, defined as twice the standard deviation of the charge distribution from its center; see the charge distributions in (b). The dashed and dotted lines are experimental phase boundaries adopted from Ref. [11]. The cyan and blue circles mark the points where a 1/4 FQHS was reported [12, 13]. These points fall in the region where the 1/2 FQHS is stabilized. The two red stars mark the points where we observe strong FQHS features at $\nu = 1/6$ (samples A and B in Table. I). Both points fall well outside the yellow region in both phase diagrams. Insets in (b) show self-consistent charge distributions (red) and potentials (black) for two samples: $n = 7.1$, $w = 58.5$ nm (bottom-left) where we observe a developing 1/6 FQHS, and $n = 26.5$, $w = 50$ nm (top-right) where a 1/4 FQHS has been reported in Ref. [12].

15

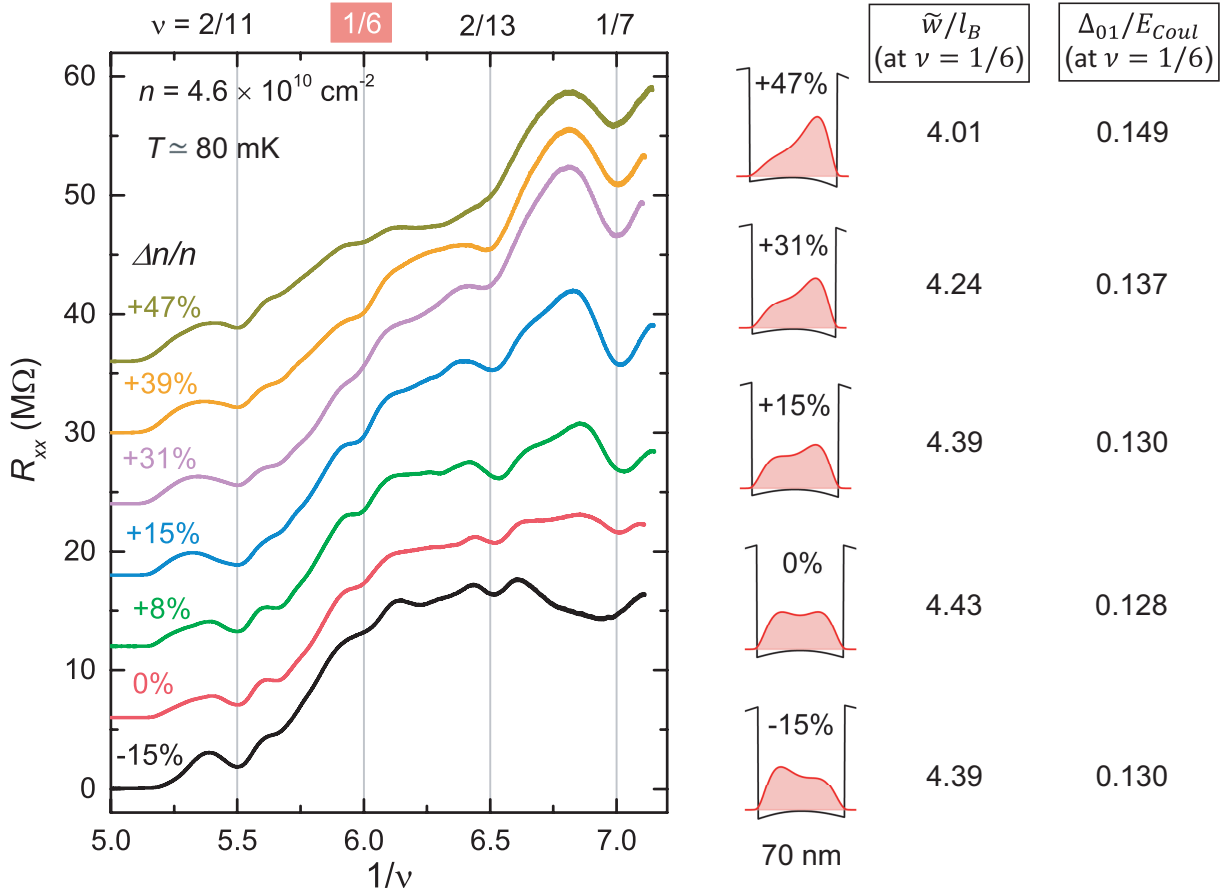

Fig. S10. **Imbalance data.** $R_{xx}$ vs $1/\nu$ traces for the 70-nm-wide GaAs QW at a fixed density of $4.6 \times 10^{10}$ cm$^{-2}$ with imbalanced charge distributions. Each trace is vertically shifted by 2 MΩ for clarity. Self-consistent charge distributions (red) and potential (black) for typical $\Delta n/n$ values are shown on the right. $\Delta n$ is defined as twice the density transferred from one side of the QW to the other. Also listed are $\widetilde{w}/l_B$ and $\Delta_{01}/E_{Coul}$, where $\Delta_{01}$ is the energy difference between the ground-state and excited-state electric subbands. For the case where the charge distribution is symmetric ($\Delta n/n = 0$), $\Delta_{01} = \Delta_{SAS}$.

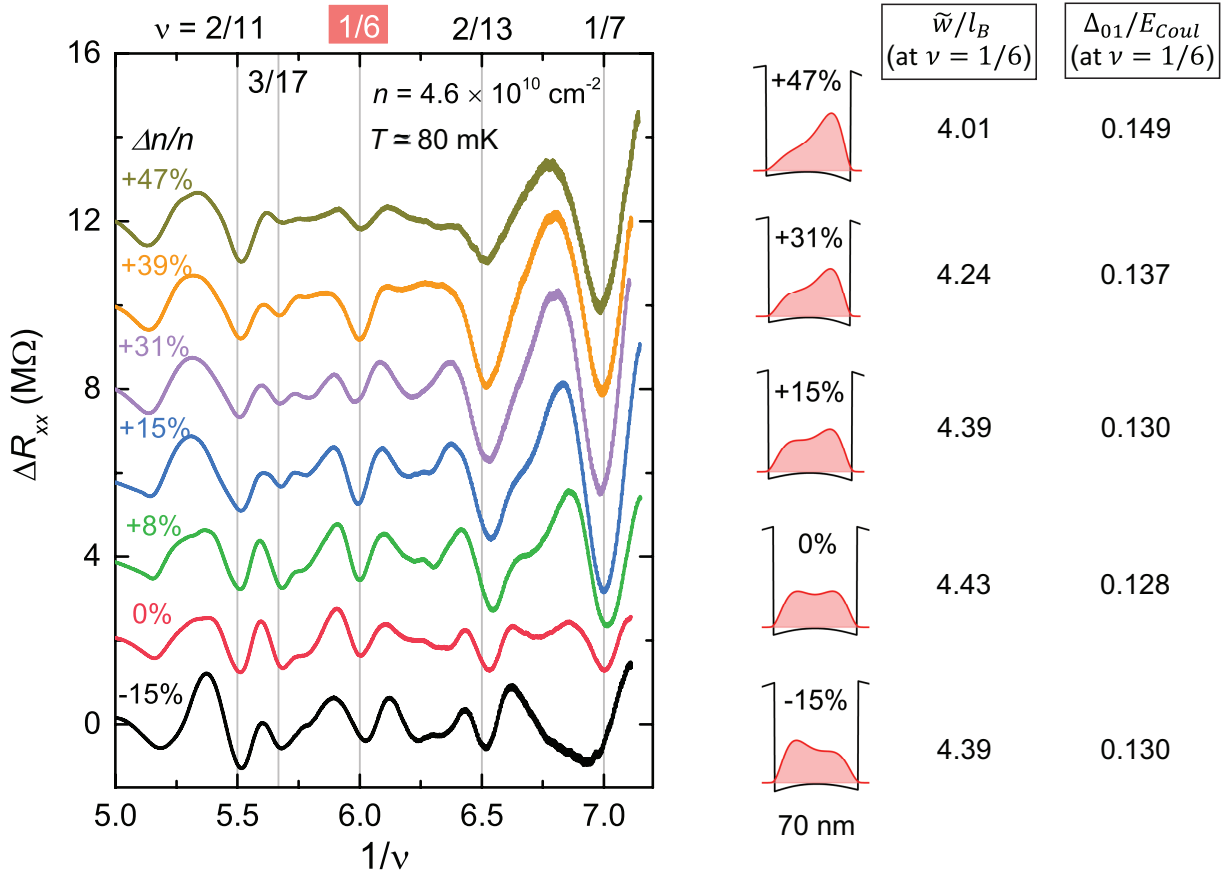

Fig. S11. **Imbalance data after subtracting smooth background.** $\Delta R_{xx}$ vs $1/\nu$ traces for the 70-nm-wide GaAs QW at a fixed density of $4.6 \times 10^{10}$ cm$^{-2}$ with imbalanced charge distributions. Each trace is vertically shifted by 2 MΩ for clarity. Self-consistent charge distributions (red) and potential (black) for typical $\Delta n/n$ values are shown on the right. $\Delta n$ is defined as twice the density transferred from one side of the QW to the other. Also listed are $\widetilde{w}/l_B$ and $\Delta_{01}/E_{Coul}$, where $\Delta_{01}$ is the energy difference between the ground-state and excited-state electric subbands. For the case where the charge distribution is symmetric ($\Delta n/n = 0$), $\Delta_{01} = \Delta_{SAS}$.

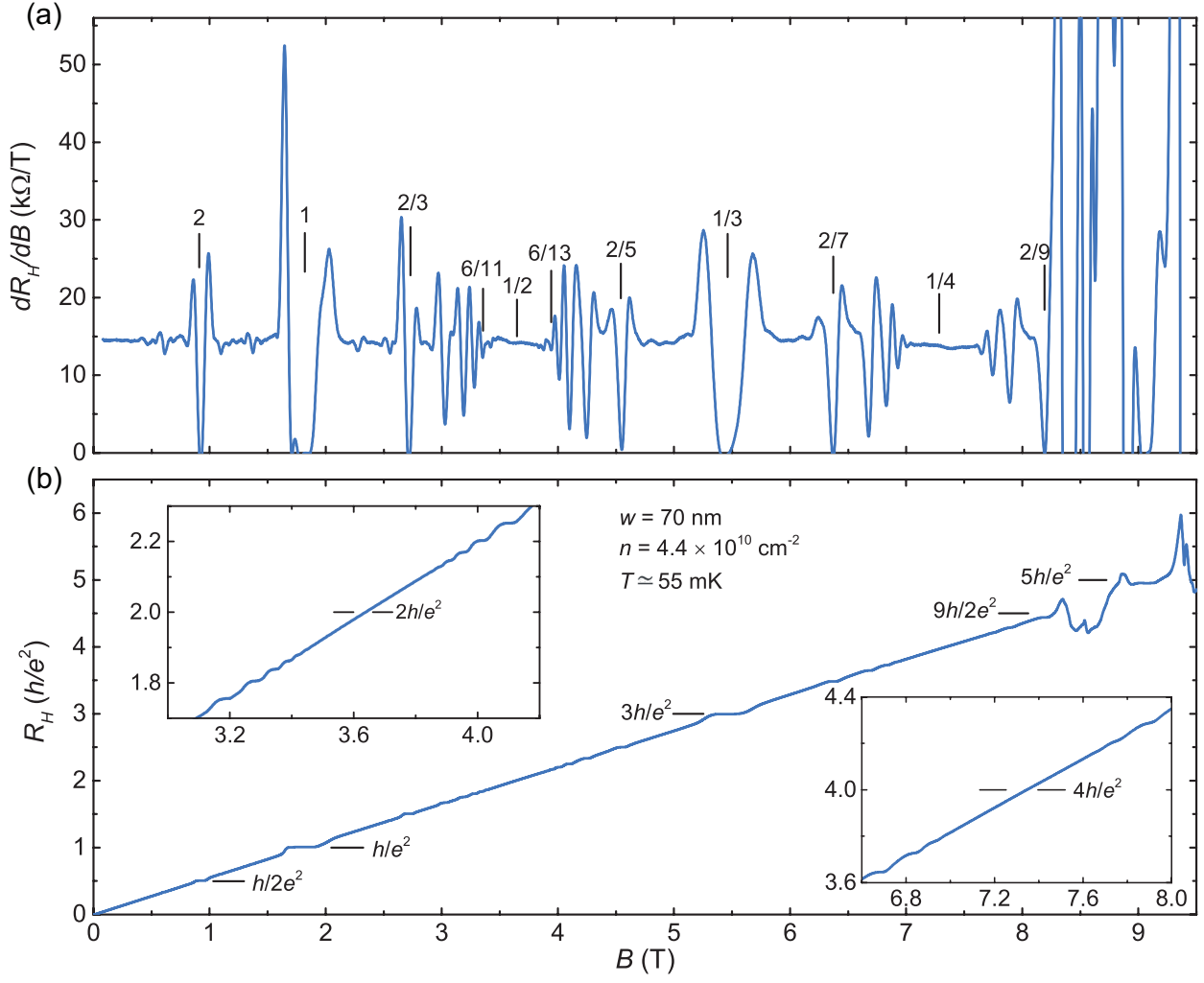

Fig. S12. **Hall data near $\nu = 1/2$ and $1/4$.** (a) Hall slope $dR_H/dB$ and (b) Hall resistance $R_H$ vs $B$ traces for sample A measured at $T \simeq 55$ mK. Insets in (b) present the enlarged version of $R_H$ vs $B$ data near $\nu = 1/2$ and $1/4$.

18

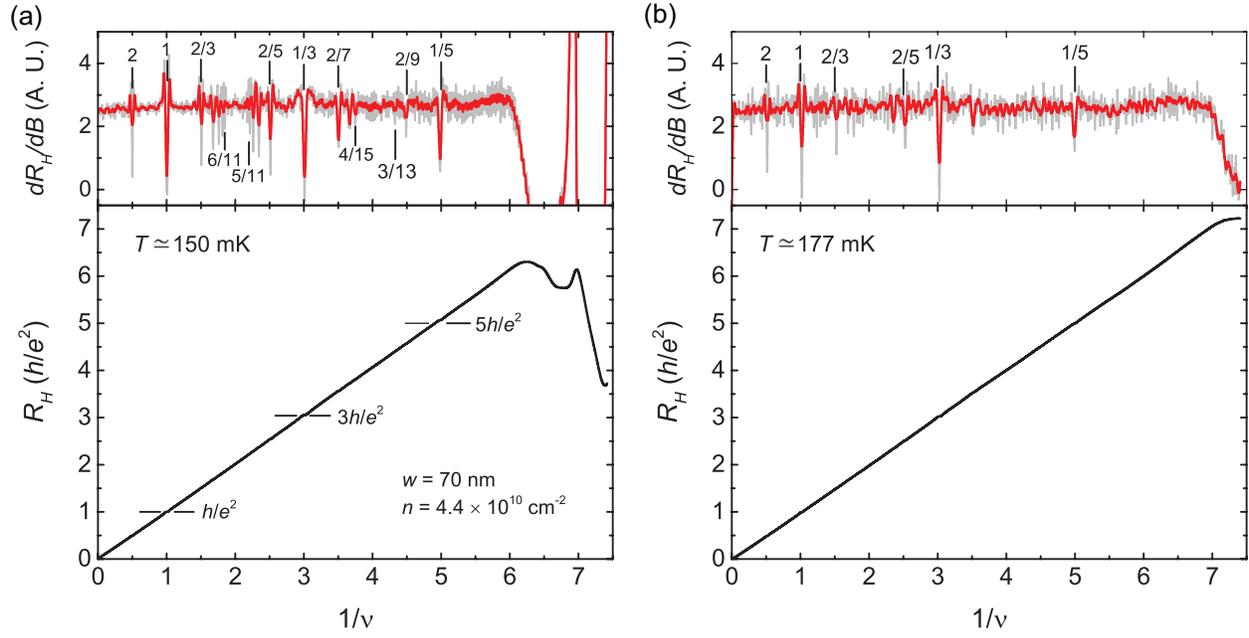

Fig. S13. **Full-field Hall data at elevated temperatures.** $dR_H/dB$ (top panels) and $R_H$ (bottom panels) vs $B$ traces for sample A measured at (a) $T \simeq 150$ mK and (b) $T \simeq 177$ mK.

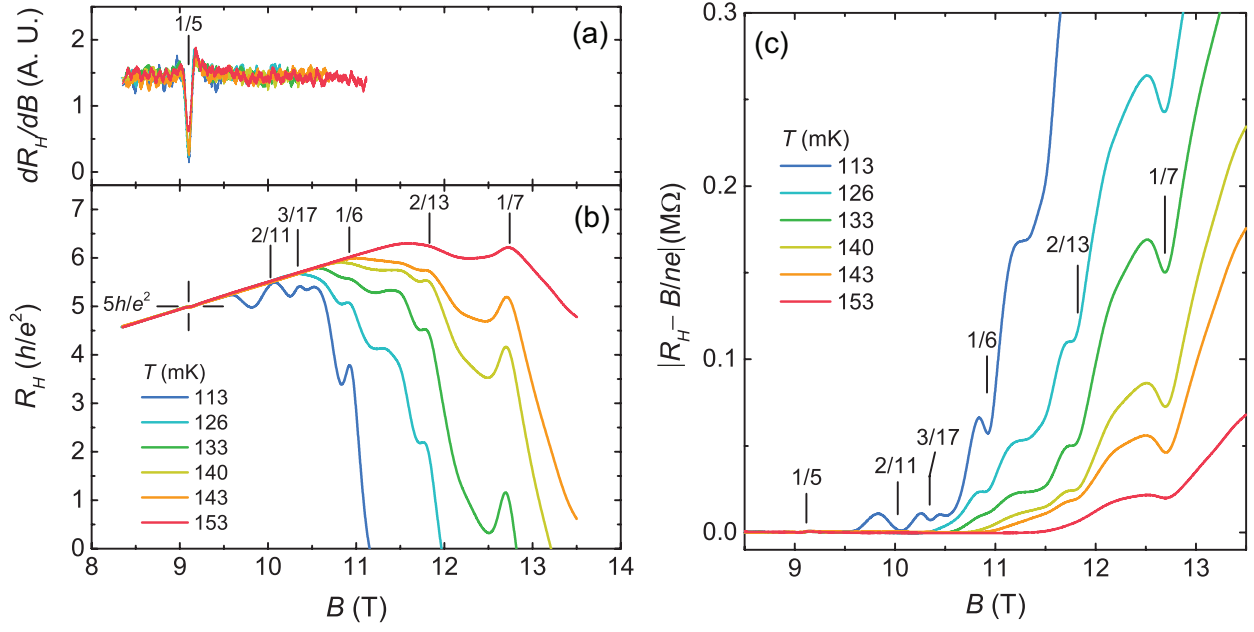

Fig. S14. **High-field Hall data at different temperatures.** (a) $dR_H/dB$ and (b) $R_H$ vs $B$ traces for sample A measured at different temperatures ranging from $T \simeq 113$ mK to $T \simeq 153$ mK. Significant $R_{xx}$ mixing effect is observed at high $B > 9$ T (low $\nu < 1/5$). (c) Deviation of measured $R_H$ from the classical Hall value $B/ne$.




\begin{thebibliography}{1}
\bibitem{Footnote.lowcurrent} In Fig. 1(a), to measure high $R_{xx}$ ($\sim$ M$\Omega$), a small driving ac current (0.2 nA) with low frequency (0.05 Hz) was used to avoid heating and nonlinear effects, and to maximize the in-phase (resistive) component of the signal.
\bibitem{Footnote.gate} This sample was fitted with front and back gates, allowing us to vary the density while also tuning the charge distribution symmetry in the GaAs QW.
\bibitem{Footnote.weaker features} We note that all the FQHS features in Fig. 2(a) are somewhat weaker than in Fig. 1(a). We attribute this to the lower 2DES quality of the gated sample because of the deposition of the metal front gate on the top surface.
\bibitem{Footnote.minimum} We note that the $B$ position of the $\nu=$1/6 $R_{xx}$ minimum in Fig. 3 is slightly lower compared to the expected value. This deviation ($\sim0.3\%$) is minor and the proximity of the minimum to $\nu=$1/6 allows us to exclude it from high-order, odd-denominator, Jain-sequence FQHSs at $\nu=p/(6p-1)$, e.g., at $\nu=$4/23 ($B=$16.8 T).
\end{thebibliography}
\end{document}